\documentclass[aps,prb,twocolumn,showpacs,floatfix,a4paper]{revtex4-2}
\usepackage{graphicx}
\usepackage[intlimits]{amsmath}
\usepackage{color}

\begin{document}

\title{Sachdev-Ye-Kitaev model: Non-self-averaging properties of the energy spectrum}
\author{Richard Berkovits}
\affiliation{Department of Physics, Jack and Pearl Resnick Institute, Bar-Ilan University, Ramat-Gan 52900, Israel}

\begin{abstract}
  The short time (large energy) behavior of the Sachdev-Ye-Kitaev model (SYK)
  is one of the main motivation to the growing interest garnered by this model. True chaotic
  behavior sets in at the Thouless time, which can be extracted from the energy spectrum.
  In order to do so, it is necessary to unfold the spectrum, i.e., to filter out global
  tendencies. Using a simple ensemble average for unfolding results in a parametically
  low estimation of the Thouless energy. By examining the behavior of the spectrum
  as the distribution of the matrix elements is changed into a log-normal distribution
  it is shown that the sample to sample level spacing variance determines this estimation
  of the Thouless energy. Using the singular value decomposition method, SVD, which filters out these
  sample to sample fluctuations, the Thouless energy becomes parametrically much larger, essentially
  of order of the band width. It is shown that 
  the SYK model in non-self-averaging even in the thermodynamic limit
  which must be taken into account in considering its short time properties.

\end{abstract}


\maketitle

\section{Introduction}

The interplay between disorder and interactions in quantum
systems has been a central theme
in condensed-matter physics for the last half century.
Recently the Sachdev-Ye-Kitaev (SYK) model
\cite{sachdev93,kitaev15}, has garnered much interest in the  fields
of quantum gravity and quantum
field theory \cite{sachdev15,maldacena16}. A key feature of the model is
that it follows random matrix theory (RMT) behavior, as is manifested in the chaotic behavior of 
its energy spectrum.

The SYK model
first appeared in the context of spin
liquids \cite{sachdev93} and then in string theory \cite{maldacena99} and
quantum gravity \cite{maldacena16}. The SYK model is known
to be chaotic \cite{garcia16,you17,li17},
showing a Wigner-like behavior of the energy spectra.
Once the SYK model is perturbed by a single-body random
term which mimics diagonal disorder in the Anderson model
\cite{garcia18,garcia18a,nosaka18,garcia19,micklitz19,monteiro21,monteiro21a},
or several SYK dots are
coupled by single-body random terms
\cite{jian17,jian17a,chen17,altland19,nandy22}, a
transition from metallic (chaotic) to insulating behavior occurs which leads
to a Wigner to Poisson crossover of the statistical properties of the spectra.

While studying nuclear and condensed-matter systems
it became clear that many physical
systems exhibits universal behavior at long times
(short energy scales)
\cite{mehta91,shklovskii93,ghur98,alhassid00,mirlin00,evers08}.
Nevertheless,
universality may break at shorter times (large
energy scales) for which a particle had no time to sample the entire
phase space of the system and its behavior depends on local non-universal
features \cite{altshuler86}.
Thus, in the context of disordered metals the
scale for which the metallic spectrum deviates from the
universal behavior is known as the Thouless energy, 
$E_{Th}=\hbar D/{\tilde L}^2=g \Delta$ ($D$ is the diffusion constant, ${\tilde L}$, is
the linear dimension, $g$ is the dimensionless conductance, $\Delta$
the average level spacing) and the
Thouless time $t_{Th}=\hbar/E_{Th}={\tilde L}^2/D$. 

The question whether an analogue of the Thouless energy manifests itself in
the SYK model has surfaced in Ref. \onlinecite{garcia16} where
Garc\'ia-Garc\'ia and Verbaarschot have 
studied (among other things) the variance of the number of levels as function
of the size  of an energy window $E$, denoted by
$\langle \delta^2 n(E) \rangle$ (where $\langle \ldots \rangle$
represents an ensemble average and $n(E)$ is the number of levels within
$E$). A departure from the RMT behavior is apparent above a certain
value of $E$, which quite naturally was identified with the Thouless energy.
Moreover, at larger energy windows,  the number variance
adopts a quadratic form $\langle \delta^2 n(E) \rangle =
a + b\langle n(E) \rangle^2$. 
Evidence for the Thouless energy has also been seen for other measures such as
off diagonal expectation values \cite{sonner17}.
In Ref. \onlinecite{altland19} the origin of the Thouless energy
for the SYK was identified as the relaxation of modes prevailing at shorter scale.
For disordered  metals these modes are know as the diffusion modes.
For SYK it was suggested that similar modes can be constructed, where the
number of such modes is connected to the number of of independent interaction
terms in the SYK model which is much smaller than the size of the
Hilbert space. This leads to an energy scale which determines the Thouless energy.
Thus, the energy scale for the departure of the SYK model level number
variance from the RMT prediction is determined
by the scale of the sample to sample fluctuations of the ensemble.

For a single disordered realization
of the SYK model all the terms have just a single variance scale,
which is well behaved (box or Gaussian) and therefore
the origin of the additional energy scale determining the Thouless energy
is not obvious. 
As noted recently by Jia and Verbaarschot \cite{jia20} this deviation from RMT
behavior has its
root in large scale sample to sample
fluctuations of the spectrum.
They attribute these fluctuations to the relatively small number of independent random
variables contributing to the SYK Hamiltonian.
When these small number of long-wavelength sample to sample
fluctuations are parameterized by terms of
Q-Hermite orthogonal polynomials and removed from the
spectrum of a particular realization
a pure RMT behavior is retained up to a very large energy scale.

The influence could be quantified \cite{jia20} by estimating
the energy scale for which the small number of independent random
variables will change the number variance. Using the notation
for the complex SYK (CSYK) half-filled model defined in the next section, where
$L$ is the number of sites, the size of the Hilbert space is
$\binom {L}{L/2} \sim 2^L$. The number of independent variables
is $\binom {2L}{4} \sim L^4$ leading to a variance of $L^{-2}$ in any
observable. Thus one expects the number of levels $n$ to deviate significantly
$O(1)$ from RMT predictions on a scale of $n \sim L^2$. A similar result
emerges from the calculation in Ref. \onlinecite{altland19} where the coefficient
$b$ in the number variance was  estimated as $b\sim L^{-4}$ thus becoming
significant at $n \sim L^2$. 

Although the deviation from RMT is shifted to larger $n$ as the system size
$L$ increases, the proportion of levels following RMT predictions
out of the the total number of states goes to zero as $L^2/2^L$.
On the other hand, one would expect that after filtering out
long wavelength sample to sample fluctuations the RMT behavior will
be followed for a finite portion of the spectrum. 
Thus, one expects the Thouless energy to crucially depend on whether one simply
averages over an ensemble or takes into account the single sample adjustments. This is
a hallmark of a non-self-averaging system \cite{aharony96}.
The number variance for the ensemble average
is different than the number variance adjusted to a particular sample.

Hence, the energy scale for which the spectrum departs from the
RMT predictions crucially depends on the the unfolding, i.e.,
the method by which the averaged
over the density of states is performed. Estimating the local density of states
by a simple ensemble average will give a different value than an
unfolding method that is able
to take into account sample specific global behavior of the spectrum.
In recent studies it has been shown that
\cite{fossion13,torres17,torres18,berkovits20,berkovits21,berkovits22,rao22},
these sample specific long ranged
features of the spectrum can be identified by the
singular value decomposition (SVD) procedure.
As detailed below, similar in a sense to Fourier transform,
SVD actually reconstructs the energy spectrum of each realization by
a sum over a series of SVD amplitudes multiplied by the corresponding SVD mode.
Unlike the Fourier transform, the amplitudes of the SVD are identical
for all realizations while the modes are realization specific. Generally,
plotting the amplitudes from large to small (known as a Scree plot) shows
that the largest amplitudes (usually $O(1)$ modes)
are orders of magnitude larger than the rest, while the following modes
amplitudes obey a
power law. The largest amplitudes modes depict the very
long wave length behavior of the
spectra, while the modes whose amplitudes follow a power law capture the
shorter range properties. Thus SVD is a very natural method to examine the SYK
spectrum behavior, and uncover
the realization specific universal properties of the spectrum.

One may conclude that there are two possible energy scales for the
deviation of the spectrum of the SYK model from RMT predictions. The
first is the energy scale corresponding to the deviation
from the ensemble average, which in self averaging systems such as
disordered metals is the Thouless energy. The second energy scale
corresponds to the deviation from RMT when the sample to sample
long-wavelength fluctuations are removed from the spectrum.
This sample specific
parameterization of these long range behavior can be done using
Q-Hermite orthogonal polynomials \cite{jia20} or
by SVD
\cite{fossion13,torres17,torres18,berkovits20,berkovits21,berkovits22,rao22},
or probably by other method.
The energy scale of these fluctuations is another relevant energy scale
which for self-averaging systems such as disordered metals is
equivalent to the Thouless energy obtained from the ensemble average
\cite{berkovits21}.
For a non-self-averaging system such as the SYK model, these two
energy scales are not equal, and for clarity we will retain the
notation of Thouless energy, $E_{Th}$, for the case where the
spectrum is unfolded by a local average over all realizations, while
the realization adjusted Thouless energy $E_{Th^*}$ is obtained
using a realization adapted unfolding method.

Here we will show that one can tweak the behavior of the
CSYK model to a more non-self-averaging behavior by changing
the distribution of the
off-diagonal to a log-normal distribution.
Thus it is possible to enhance the sample to sample fluctuations
and study its influence on $E_{Th}$ and $E_{Th^*}$.
Such a wide distribution was not previously considered for the SYK
model and should help to clarify the divergence of $E_{Th^*}$ from $E_{Th}$.

The paper is arranged as follows: CSYK is defined in
Sec. \ref{s1}. Corroborating the expected behavior for short range energy
scales is performed in Sec. \ref{s2}. The universal statistics
4-fold symmetry as function of the system size is observed.
In Sec. \ref{s3} long range energy scales are probed by the 
number variance. The spectrum is unfolded by using the local ensemble averaged
level spacing. RMT predictions hold only up to $E_{Th}$, which
becomes smaller as the log-normal distribution acquires a thicker tail towards
larger values. Above $E_{Th}$ the variance increases quadratically
as function of the number of levels in the energy window. In Sec. \ref{s4}
we switch
to the SVD analysis. This analysis reveals that each realization has
a distinct long-range correlated level spacing structure. Thus, one
should adapt the unfolding for each realization. After a realization
adapted unfolding the number variation follows the RMT prediction for
much larger energy scales, i.e., $E_{Th^*} \gg E_{Th}$. Actually, $E_{Th}$
could be estimated from the sample to sample variance in the level spacing.
In Sec. \ref{s5} these results are discussed in the limit of large CSYK
systems, showing the in the thermodynamic limit the SYK is non-self-averaging.

\section{Complex SYK model}
\label{s1}

Here we use the the complex spinless fermions version of the
SYK model given by the following Hamiltonian:
\begin{eqnarray} \label{syk}
  \hat  H=
  \sum_{i>j>k>l}^{L} V_{i,j,k,l} \hat c_i^\dag \hat c_j^\dag \hat c_k \hat c_l,
\end{eqnarray}
the couplings $V_{i,j,k,l}$ are complex numbers, where the
real and imaginary components are  independently drawn
from an identical distribution. We study here two different distribution:
The first is a box distribution between $-L^{-3/2}/2 \ldots L^{-3/2}/2$,
where L is the number of sites. The second distribution
is the log-normal distribution
\begin{eqnarray} \label{syk1}
P(V) =(A/|V|) e^{\frac {-\ln^{2}  (|V|/V_{\rm typ})} {2p \ln (V_{\rm typ}^{-1})}},
\end{eqnarray}
with $V_{\rm typ} \sim K^{-\gamma/2}$, $K$ is the size of the Hilbert space, 
$A$ a normalization and
$\gamma$, $p$ are parameters. For simplicity here we shall
set $p=0.5$, while $\gamma$ is varied \cite{khaymovich20}.
Thus, the log-normal  distribution becomes wider
and more skewed as $\gamma$ increases.
The number of fermions is conserved  and we considered the $N=L/2$
sector for even $L$ and $N=(L+1)/2$ sector for odd $L$.
Resulting in a Hilbert space size of $K=\binom {L}{N}$, and
a matrix size of $K \times K$.

\section{Short Energy Scales}
\label{s2}

As a first step we would like to probe the nearest neighbor level spacing
statistics of the  CSYK box distributed in order to
establish the extended regime of this model. One expects
that in the extended regime short energy scales (long times) follow the
Wigner statistics. The short energy scale statistics is revealed by the
ratio statistics, defined as:
\begin{eqnarray} \label{ratio}
r &=& \langle \min (r_n,r_n^{-1}) \rangle,
\\ \nonumber
r_n &=& \frac {E_n-E_{n-1}}{E_{n+1}-E_{n}},
\end{eqnarray}
where $E_n$ is the n-th eigenvalue of the Hamiltonian
and $\langle \ldots \rangle$ is
an average over an ensemble of different realizations of disorder and half of the eigenvalues
around the middle of the band.
For the Wigner  distribution $r_s \cong  0.5307$
for the GOE symmetry, $r_s \cong  0.5996$ for GUE and $r_s \cong 0.6744$ for GSE.
\cite{atas13}.

\begin{figure}
\includegraphics[width=10cm,height=!]{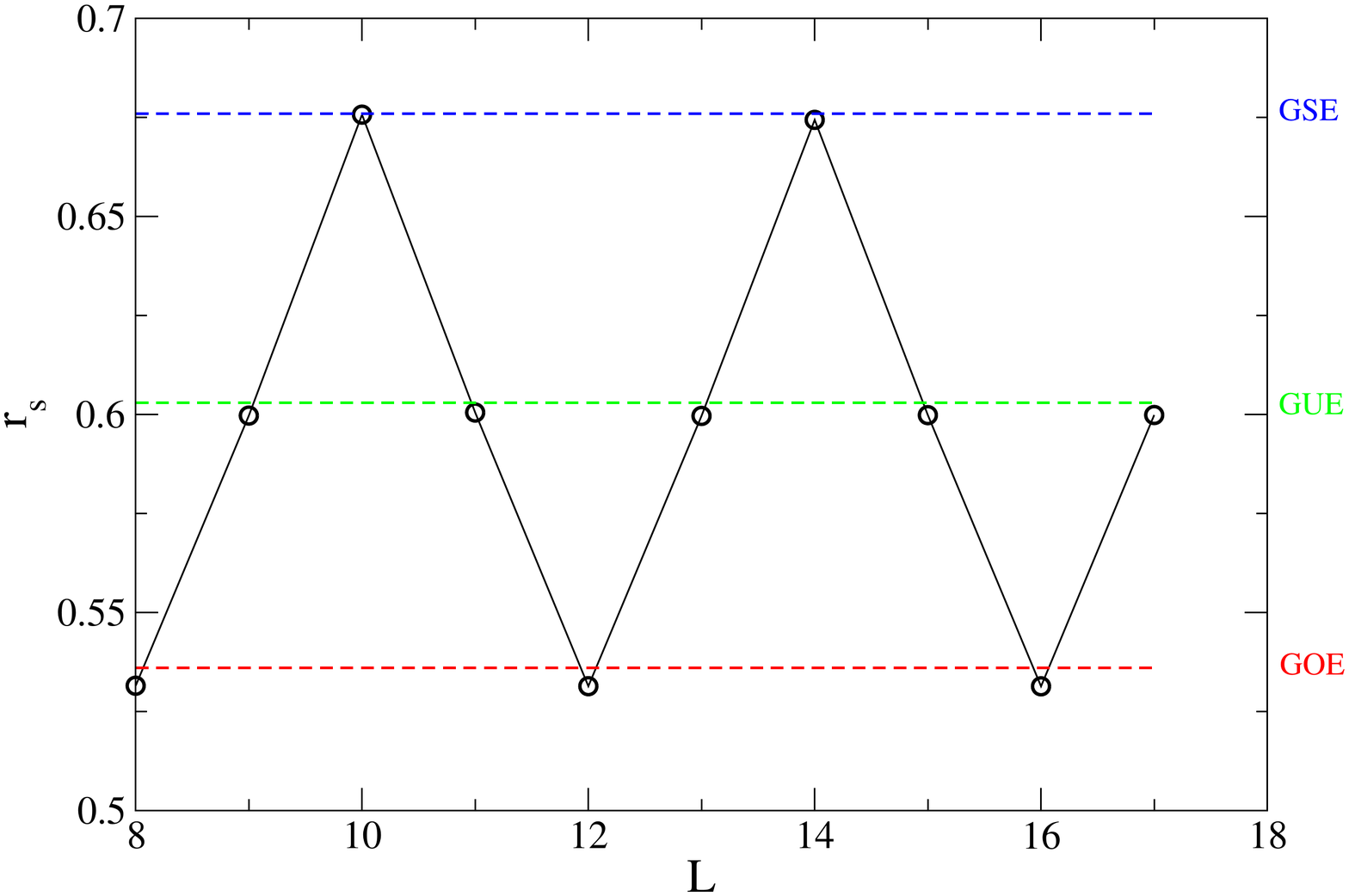}
\caption{\label{fig1}
  The nearest neighbor level spacing statistics as manifested in the
  behavior of the ratio statistics $r_s$ for different system sizes $L$
  of the CSYK with a box distribution are indicated by the black
  circles.
  The number of fermions is $N=L/2$ for even $L$ and $N=(L+1)/2$ for odd $L$.
  The $r_s$ values expected for GOE (dashed red) GUE (dashed green) and
  GSE (dashed blue) are marked. The expected 4-fold symmetry is seen.
}
\end{figure}

An interesting behavior emerges for the CSYK. It is
known that as a consequence of the symmetries of the SYK model, the
spectrum of Eq. (\ref{syk})
shows statistics which depends on $L$ 
\cite{garcia16,you17,li17}. For $L \mod 4 =0$ the statistics are GOE,
For $L \mod 4 =2$ the statistics are GSE, and for $L \mod 4 = 1,3$ the
statistics are GUE. This is indeed seen in
Fig. \ref{fig1} where $r_s$ averaged over
the middle half of the energy spectrum,
for different sizes
$L=8,9,10,11,12,13,14,15,16,17$ and number of fermions
$N=4,5,5,6,6,7,7,8,8,9$ is plotted. In all cases we exactly diagonalize the
corresponding $K \times K$ matrix and average over $3000$ realizations of
disorder (except for the largest size for which only $100$  realization were
computed). It can be seen that the expected 4-fold symmetry is followed.

Concentrating on the $L=16$ with $N=8$ systems, we investigate the role
played by changing the distribution from the box distribution to a
log-normal. Setting $p=0.5$ and increasing $\gamma$ we sweep through
the values $\gamma=1,1.5,2,2.5,3,4,5$. For all these values
$r_s \cong  0.5307 \pm 0.001$. Thus, GOE universal behavior on short
energy scales is perfectly followed.


\section{Local ensemble unfolding }
\label{s3}

As discussed, the practice of determining 
the Thouless energy is fraught with difficulties. In order 
to compare any spectrum to RMT predictions, one must
recast the spectrum such that it will exhibit an averaged
constant density of states, i.e., a constant level spacing throughout
the region examined. What is
the averaging procedure? Usually, one averages the level spacing over an
ensemble of disordered realizations in a given region and then reconstructs a particular
spectrum such as the level spacing is on the average equal to 1 everywhere.
Specifically, the averaged level spacing for the i-th level
is $\Delta_i = \langle E_{i+m}-E_{i-m}\rangle/2m$ (where $m$ is $O(1)$, here chosen
as $m=5$), and the unfolded spectrum for the j-th realization is
$\varepsilon^j_i=\varepsilon^j_{i-1}+(E^j_i-E^j_{i-1})/\Delta_i$.
For brevity we shall call this
unfolding procedure local ensemble unfolding.

\begin{figure}
\includegraphics[width=10cm,height=!]{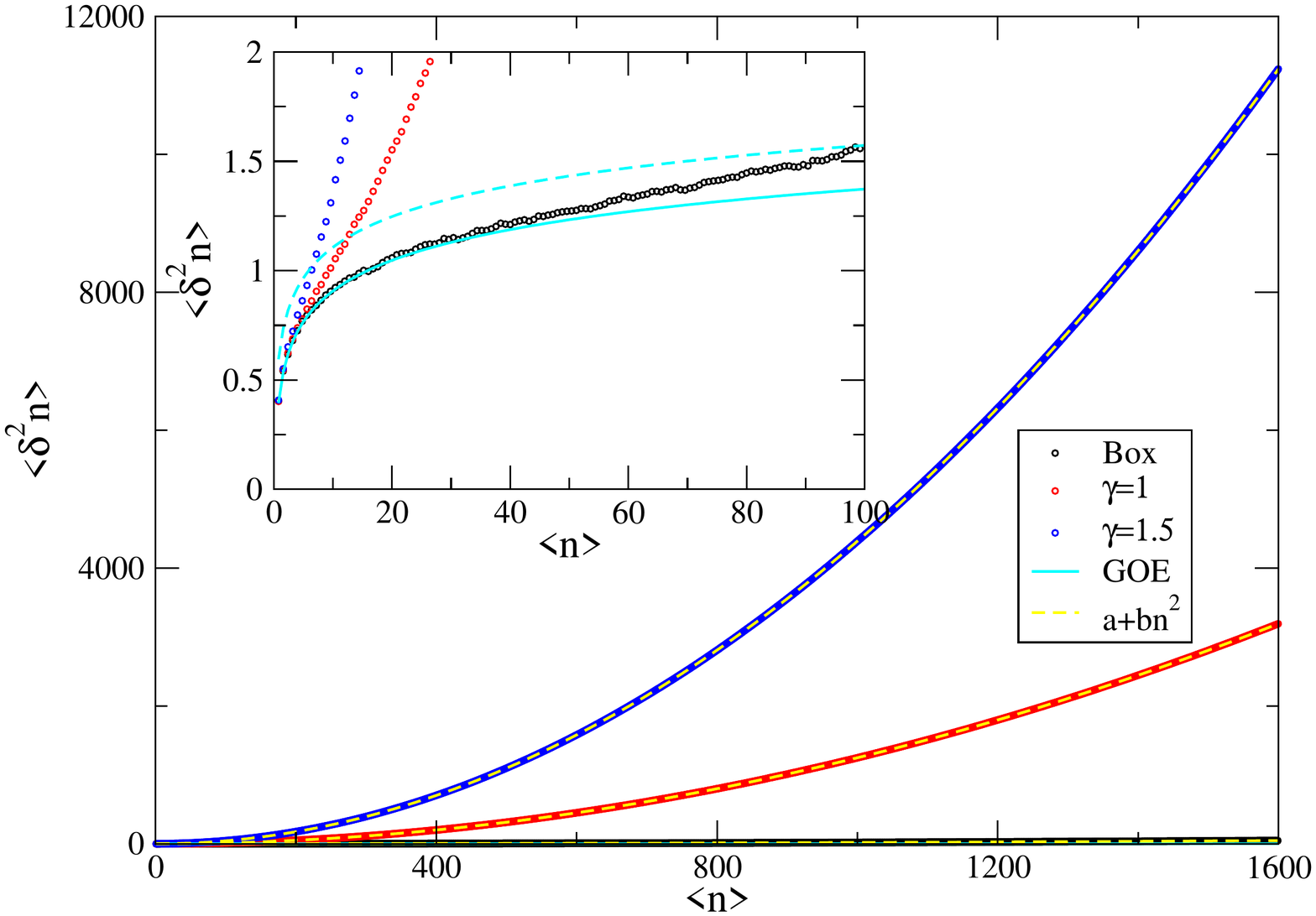}
\caption{\label{fig2}
  The level number variance, $\langle \delta^2 n(E) \rangle$, as
  function of the energy window $E$. Symbols (black - box, red - $\gamma=1$,
  blue $\gamma=1.5$ represent the variance
  with local ensemble unfolding, fitted for larger values by $a+b\langle n(E) \rangle^2$
  (where $b=1.78 \ 10^{-5}$, for the box distribution, $b=1.25 \ 10^{-3}$, for
  $\gamma=1$, and $b=4.4 \ 10^{-3}$, for $\gamma=1.5$. The cyan
  line is the GOE prediction $\langle \delta^2 n(E) \rangle =
  (2/\pi^2) \ln(\langle n(E) \rangle) + 0.44$.
  Insert: zoom into smaller values of $<n>$. Deviation from the GOE behavior
  are observed. A curve depicting GOE plus a constant of $0.2$ corresponds
  the dashed cyan line. Using the intersection between the variance and
  the dashed cyan line to determine the Thouless energy results in
  $E_{Th} \sim 95 \Delta$ for the box distribution,
  $E_{Th} \sim 13 \Delta$, for $\gamma=1$, and  $E_{Th} \sim 7 \Delta$
  for $\gamma=1.5$.
 }
\end{figure}

Implementation of this local unfolding procedure for the level number variance
of CSYK with a box distribution  and for a log-normal distribution
$\gamma=1,1.5$
results in the behavior depicted in the inset of
Fig \ref{fig2}, which is in agreement
with the behavior observed in Refs. \cite{garcia16,jia20}.
Here the number variance begins by following the GOE predictions
and grows quadratically for larger $\langle n \rangle$.
The Thouless energy corresponds to the point where the number variance
deviates from GOE predictions. As an estimate of the Thouless energy
we chose the point for which the variance deviates by an arbitrary amount
(set as $0.2$), resulting in
$E_{Th} \sim 95 \Delta$, for the box distribution,
$E_{Th} \sim 13 \Delta$, for $\gamma=1$, and  $E_{Th} \sim 7 \Delta$
for $\gamma=1.5$, where $\Delta=\langle \Delta_i \rangle$.
This will naturally lead to the conclusion that
as $\gamma$ increases, $E_{Th}$ strongly decreases.
For larger values of $\gamma$ the variance departs from the GOE predictions
close to $\Delta$ and were not plotted to avoid cluttering the figure at small
$\langle n \rangle$.

As emphasized by Jia and Verbaarschot \cite{jia20}, since unlike typical
RMT models for which all non-diagonal terms are are random, SYK models have
just $\binom {2L}{4} \ll \binom {L}{N}$ independent
non-diagonal terms, which leads to significant sample to sample fluctuations
within the ensemble and using the average level spacing obtained by ensemble
average may significantly skew the number variance. Thus, we need a way to
characterize the level spacing for a specific sample.
As previously  discussed, in Ref. \onlinecite{jia20}
this was achieved by parameterized the  spectrum using
Q-Hermite orthogonal polynomials. We will use the SVD method as described
in the next section.

\section{Singular Value Decomposition}
\label {s4}

SVD can be used to characterize the features of the spectrum, for example
on what scale does it follow RMT predictions
\cite{fossion13,torres17,torres18,berkovits20,berkovits21,berkovits22,rao22}.
For the analysis, one tabulates $P$ eigenvalues around the center of the band
of $M$ realizations of disorder
as a matrix $X$ of size $M \times P$ where $X_{mp}$ is the $p$ level of
the $m$-th realization. The matrix
$X$ is decomposed to $X=U \Sigma V^T$, where
$U$ and $V$  are $M\times M$ and $P \times P$ matrices correspondingly,
and $\Sigma$ is a {\it diagonal} matrix of size $M \times P$ and rank
$r=\min(M,P)$. The $r$ diagonal elements of $\Sigma$
are the singular values amplitudes (SV) $\sigma_k$ of $X$.
All $\sigma_k$ are positive
and therefore may be ordered by their size
$\sigma_1 \geq \sigma_2 \geq \ldots \sigma_r$.
The Hilbert-Schmidt norm of the matrix
$||X||_{HS}=\sqrt{Tr X^{\dag}X}=\sum_k \lambda_k$ (where $\lambda_k=\sigma_k^2$).
The matrix $X$ could be written as a series composed of matrices $X^{(k)}$,
where $X^{(k)}_{ij}=U_{ik}V^T_{jk}$ and $X_{ij}=\sum_k \sigma_k X^{(k)}_{ij}$.
This series in an approximation
of matrix $X$, where the sum of the first $m$ modes gives a matrix
$\tilde X =\sum_{k=1}^m \sigma_k X^{(k)}$, for which $||X||_{HS}-||\tilde X||_{HS}$
is minimal. 

The idea is that low modes capture the 
long-wavelength fluctuations of the spectrum, while higher modes sample
the shorter wavelength fluctuations. If a distinct pattern of behavior of the
amplitude as function of the mode number can be seen for a particular
range of $k$, it is meaningful to
discuss different behaviors at different energy scales. Indeed,
examining 
the scree plot of the singular values $\lambda_k$
for box distribution and log-normal distribution with different
values of $\gamma$ and a fixed $p=0.5$,
for $L=16$, $N=8$, with $M=4096$ realizations and $P=4096$
eigenvalues around the middle of the band one sees two distinct regions.
The lowest modes ($k=1,2$ for box and $\gamma=1$; $k=1,2,3,4$ for
$1.5\leq \gamma \leq 2.5$; and $k=1,2,3,4,5,6$ for $3 \leq \gamma \leq 5$)
amplitudes
are much larger than all the other modes, and
determine the very long-wave behavior of the spectrum.
The bulk of the modes for larger $k$ follow a power-law behavior
($\lambda_k \sim k^{-\alpha}$) with $\alpha=1$, as expected for
Wigner-Dyson statistics \cite{fossion13,torres17,torres18}.

\begin{figure}
\includegraphics[width=10cm,height=!]{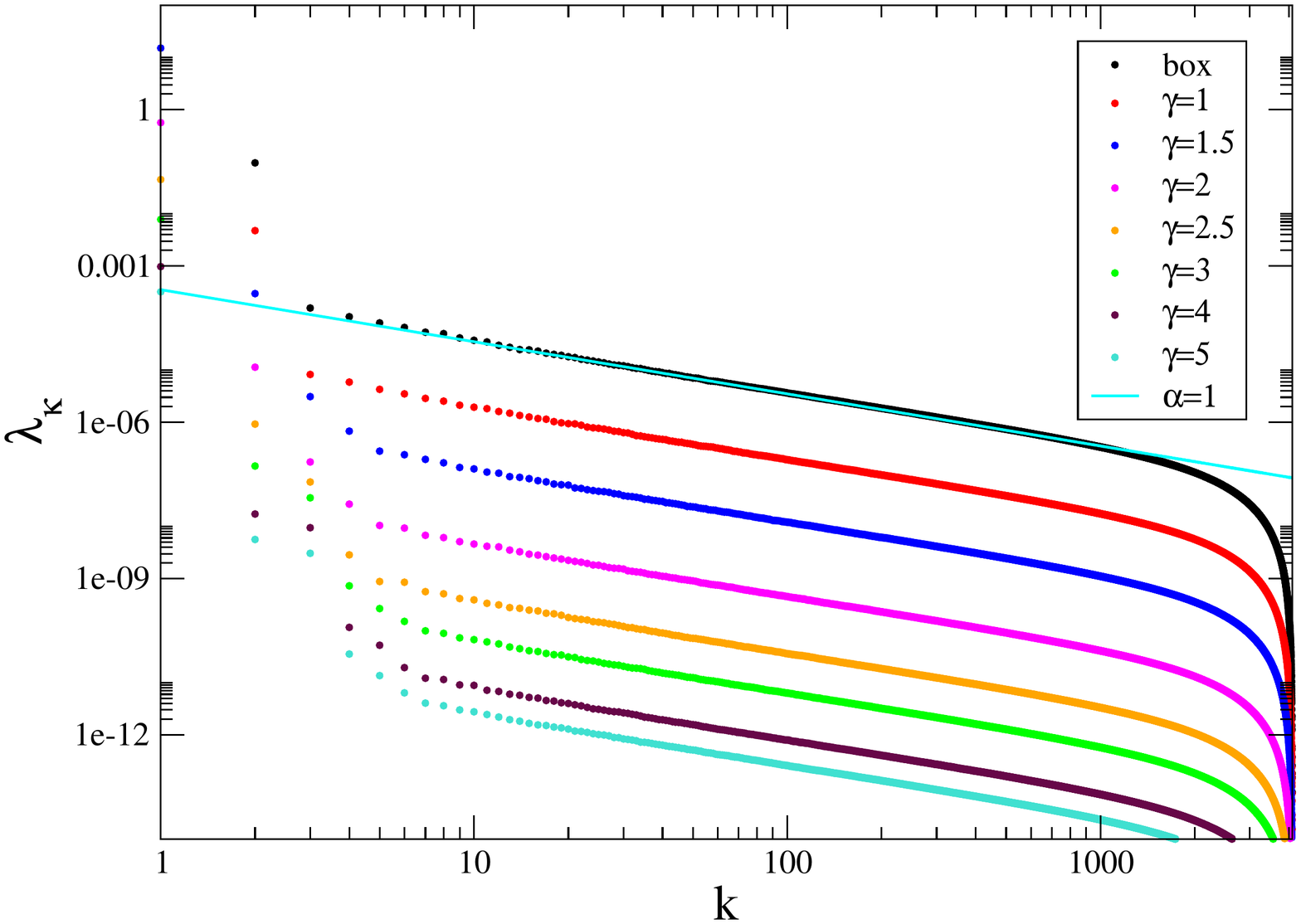}
\caption{\label{fig3}
  The scree plot of the singular values for the CSYK with box
  and log-normal distribution with $p=0.5$ and $\gamma=1,1.5,2,2.5,3,4,5$
  for $L=16$, $N=8$, with
  $M=4096$ realizations and $P=4096$ eigenvalues around the middle of the band.
  The square amplitude of the singular vale $\lambda_k$ are indicated by
  the symbols. The cyan line corresponds to
  $\lambda_k \sim k^{-\alpha}$, with $\alpha=1$, as expected for a spectrum
  which follows Wigner-Dyson statistics. The lower $k$ modes
  which capture the non-universal global structure of the spectrum deviate
  from Wigner-Dyson.
}
\end{figure}

\begin{figure}
\includegraphics[width=10cm,height=!]{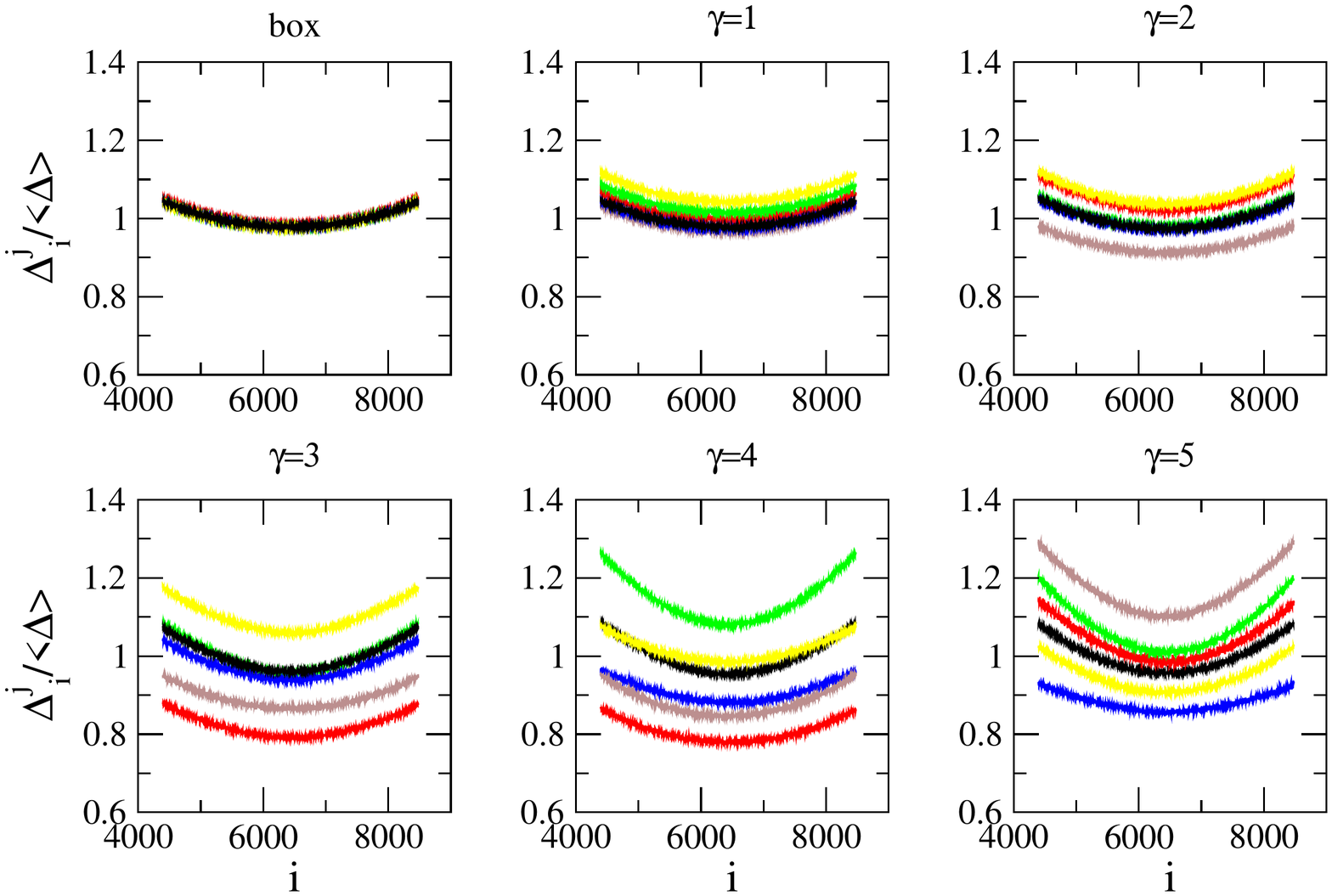}
\caption{\label{fig4}
  The level spacing for the CSYK with box distribution and different
  log-normal distribution with $\gamma=1,2,3,4,5$ for $L=16$, $N=8$, for
  $M=4096$ realizations and $P=4096$ eigenvalues
  around the middle of the band. The ensemble
  averaged level spacing, $\Delta_i$, for
  the local unfolding method is the same for all samples and is depicted
  by the black curve. The SVD unfolded level spacing
  , $\Delta^j_i$, is realization dependent.
  Five individual realizations for each distribution are shown
  (red,green,blue,yellow,brown curves). As $\gamma$ increases the
  sample to sample fluctuations increases. It is clear that the long range
  correlations within a sample are very significant.
}
\end{figure}

In order to illustrate the difference between the local
ensemble unfolding and unfolding
using the lower modes  of the SVD it is useful to examine the difference
in the behavior of level spacing of the i-th level, $\Delta_i$
obtained by each method.
While for the local ensemble unfolding the level spacing $\Delta_i$ is
averaged over all realizations and therefore is not realization dependent,
for the SVD unfolding the i-th level spacing of the j-th realization
$\Delta^j_i = (\tilde \varepsilon^j_{i+m}-\tilde \varepsilon^j_{i-m})/2m$, where
$\tilde \varepsilon^j_i=\sum_{k=1}^2 \sigma_k U_{ik}V^T_{jk}$,
is realization dependent. 
As can be seen in Fig. \ref{fig4} there is a noticeable difference
between the realization specific level spacing $\Delta^j_i$ and the
ensemble average $\Delta_i$. This difference becomes larger as
$\gamma$ increases, i.e.. as the width of the log-normal distribution increases.
Moreover, the difference, $\Delta^j_i-\Delta_i$,
is long-range correlated for a given realization across thousands of levels.
Thus, it makes sense to define the typical spacing difference between
realizations
$\delta \Delta = \sqrt{\langle (\Delta^j_i-\Delta_i)^2\rangle_i}$.
For $L=16$, $N=8$ with $M=4096$ realizations and $P=4096$ levels around
the center of the band one gets $\delta \Delta = 4.44 \ 10^{-3}
\Delta$ for the box distribution
$\delta \Delta = 3.52 \ 10^{-2} \Delta$ for $\gamma=1$
and $\delta \Delta = 6.61 \ 10^{-2} \Delta$ for $\gamma=1.5$.

This long-range sample to sample level fluctuations can explain the behavior
of the level number variance of the local ensemble unfolding seen in
Fig. \ref{fig3}. Essentially, for larger values of $\langle n \rangle$,
$\langle \delta^2 n \rangle = a+b \langle n \rangle^2$ with
$b=1.78 \ 10^{-5}$, for the box distribution, $b=1.25 \ 10^{-3}$, for
$\gamma=1$, and $b=4.4 \ 10^{-3}$, for $\gamma=1.5$. 
The quadratic behavior could be understood as the
consequence of realization specific long-range fluctuation of the level
spacing. 
Calculating $\langle \delta^2 n \rangle =
\langle (n -\langle n \rangle)^2 \rangle$, taking into account
that after unfolding $\langle n \rangle = n$ and for a typical
realization $n \sim n(1+\delta \Delta /\Delta)$.
Thus, $\langle \delta^2 n \rangle \sim
(n \delta \Delta/ \Delta)^2$, and after plugging in the
above mentioned values of $\delta \Delta $ 
one obtains $\langle \delta^2 n \rangle \sim 1.97 \ 10^{-5} n^2$
for the box distribution,
$\langle \delta^2 n \rangle \sim 1.24 \ 10^{-3} n^2$ for $\gamma =1$ and
$\langle \delta^2 n \rangle \sim 4.36 \ 10^{-5} n^2$ for $\gamma =1.5$
in good
agreement with the numerical value quoted above.
Moreover, using our
previous definition of the Thouless energy as the energy for which
the deviation from RMT results becomes larger than some threshold,
$b (E_{TH}/ \Delta)^2 =0.2$, and using
$b=(\delta \Delta /\Delta)^2$, one gets
$E_{Th}= 0.44 \Delta^2/\delta \Delta$ resulting in
$E_{Th}= 100 \Delta,18.5 \Delta,6.7 \Delta$ for box and log-normal
distributions with $\gamma=1,1.5$ correspondingly, in good agreement
with the results in Fig. \ref{fig3}.

One concludes that the main contribution to the local ensemble
average Thouless energy, $E_{Th}$,
comes from the sample to sample long range fluctuations
which can be quantified by $\delta \Delta$,
the typical level spacing difference between the different realizations
in the ensemble. 

\begin{figure}
\includegraphics[width=10cm,height=!]{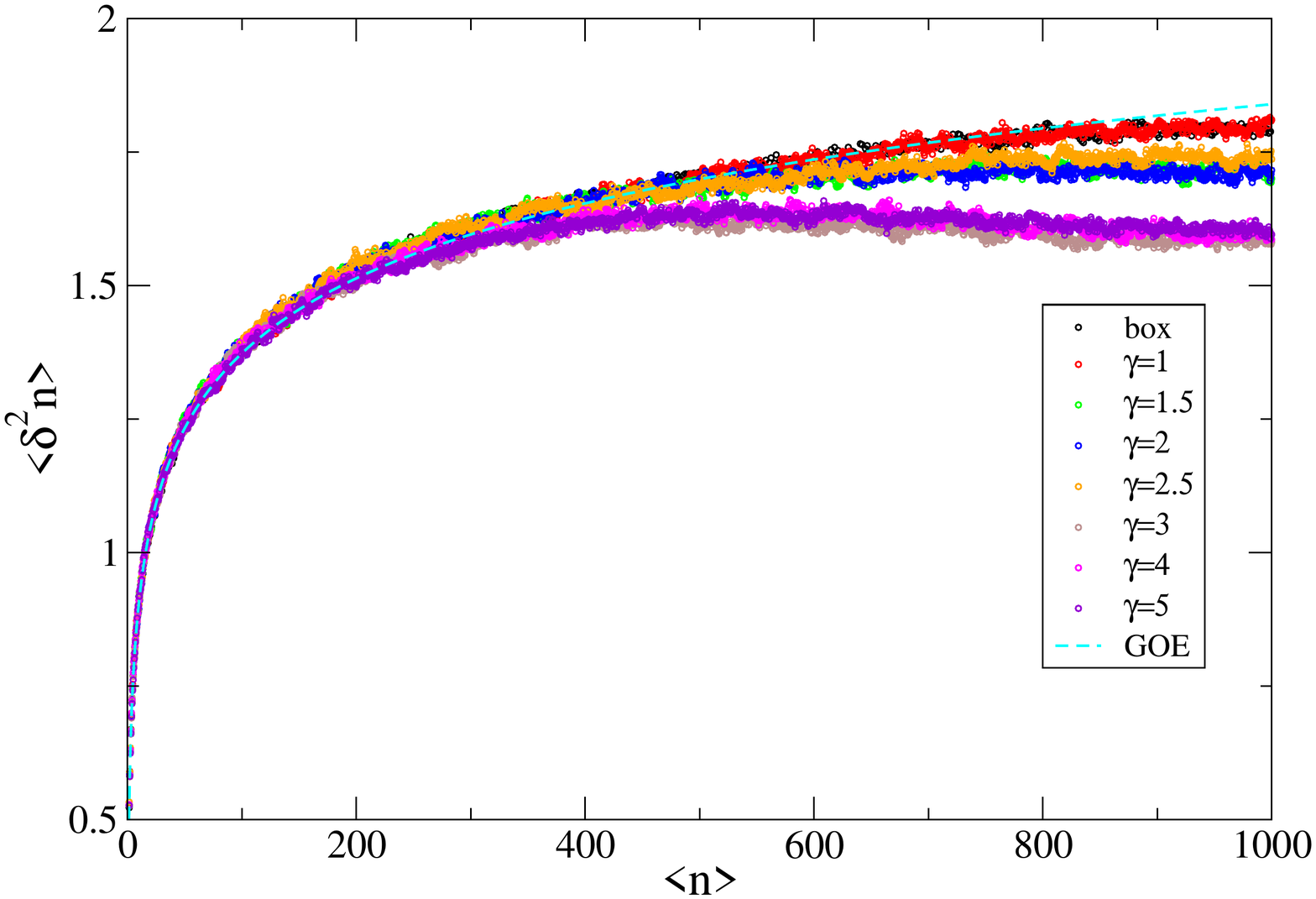}
\caption{\label{fig5}
  The SVD unfolded 
  level number variance, $\langle \delta^2 n(E) \rangle$, as
  function of the energy window $E$
  for box and log-normal
  $\gamma=1,1.5,2,2.5,3,4,5$ represented by the symbols.
  The cyan dashed lines
  line is the GOE prediction $\langle \delta^2 n(E) \rangle =
  (2/\pi^2) \ln(\langle n(E) \rangle) + 0.44$.
  The SVD unfolded number variance departs from GOE predictions
  at $E_{Th^*}\sim 800\Delta$ for the box and $\gamma=1$ distributions,
  at $E_{Th^*}\sim 500\Delta$ for $1.5 \leq \gamma \leq 2.5$ distributions,
  and at $E_{Th^*}\sim 300\Delta$ for $3 \leq \gamma \leq 5$ distributions.
}
\end{figure}

Thus,
using the sample specific level spacing from the SVD (i.e., $\Delta^j_i$)
for unfolding will eliminate the sample to sample fluctuations
contribution to the number variance. Indeed, using the
SVD unfolded spectrum of the $j$-th realization $\tilde \varepsilon^j_i$
defined by 
$\varepsilon^j_i=\varepsilon^j_{i-1}+
(\tilde \varepsilon^j_i-\tilde \varepsilon^j_{i-1})/\Delta^j_i$,
for the calculation of the realization specific level number variance
one obtains
the values depicted in Fig. \ref{fig5}. The number
variance fits well with GOE expectations up to
$E_{Th^*}\sim 800\Delta$ for the box and $\gamma=1$ distributions,
$E_{Th^*}\sim 500\Delta$ for $1.5 \leq \gamma \leq 2.5$ distributions,
and $E_{Th^*}\sim 300\Delta$ for $3 \leq \gamma \leq 5$ distributions.
The deviation is towards lower variance than predicted by GOE, similar
to the behavior of the number variance after unfolding using
Q-Hermite orthogonal polynomials \cite{jia20}. 
As shown in Ref. \onlinecite{berkovits22} the energy of deviation
from  the universal behavior can be read off the scree plot by defining
the mode for which the behavior deviates from the $k^{-1}$ power law as
$k_{Th^*}$. Thus, as can be garnered from Fig. \ref{fig3},
$k_{Th^*}\sim 2$ for the box and $\gamma=1$ distributions,
$k_{Th^*}\sim 4$ for $1.5 \leq \gamma \leq 2.5$ distributions,
and $k_{Th^*}\sim 6$ for $3 \leq \gamma \leq 5$ distributions.
Estimating the Thouless energy by $E_{Th^*}=r \Delta/2 k_{Th^*}$ \cite{berkovits22}
results in $E_{Th^*}\sim 1000\Delta$, $E_{Th^*}\sim 500\Delta$, and
$E_{Th^*}\sim 300\Delta$, not to far from the estimations
obtained from Fig. \ref{fig5}.

\begin{figure}
\includegraphics[width=10cm,height=!]{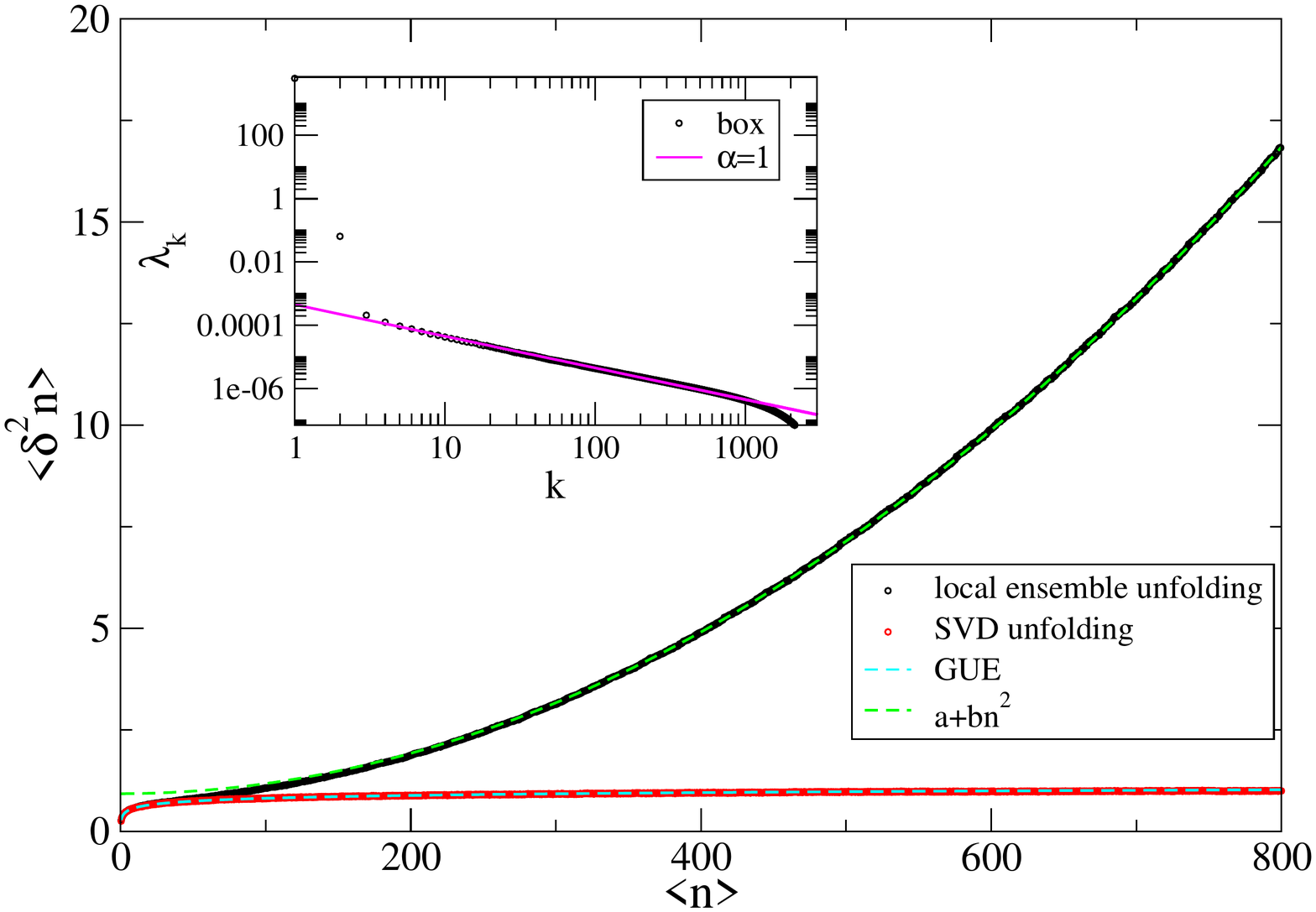}
\caption{\label{fig6}
  The level number variance, $\langle \delta^2 n(E) \rangle$, as
  function of the energy window $E$ for box distribution
  using local ensemble unfolding (black symbols) and SVD unfolding
  (red symbols) for $L=15$, $N=8$, $P=3000$ and $M=3000$.
  The cyan dashed lines
  line is the GUE prediction $\langle \delta^2 n(E) \rangle =
  (1/\pi^2) \ln(\langle n(E) \rangle) + 0.35$, while green dashed
  line follows $a+b n^2$  with $b=2.48 \ 10^{-5}$.
  The different behavior for the two unfolding methods is clear.
  Inset: the scree plot of the singular vale $\lambda_k$  indicated by
  the symbols as function of $k$ . The purple line corresponds to
  $\lambda_k \sim k^{-\alpha}$, with $\alpha=1$.
}
\end{figure}

Up to now we have mainly presented results for $L=16$ which
is the largest system for which we have ample statistics. Nevertheless,
as can be seen in Fig. \ref{fig1}, for the CSYK the statistics changes for
different values of $L$. In order to verify that our conclusions regarding
the non-self-averaging behavior of the model are relevant for any symmetry,
we calculated $\langle \delta^2 n(E) \rangle$ for the box distribution
using both local ensemble unfolding and SVD unfolding
for $L=15$, with $N=8$ particles, for $P=3000$ eigenvalues around
the middle of the band for $M=3000$ realizations. For this case GUE
behavior, i.e., $\langle \delta^2 n(E) \rangle = (1/\pi^2) \ln(\langle n(E)
\rangle) + 0.35$. Indeed, as can be seen in Fig. \ref{fig6}, both unfolding
methods fit the GUE expectation  for small $\langle n \rangle$. On
larger scales the same difference that was seen for $L=16$ GOE is seen
for $L=15$ GUE systems. The local ensemble unfolding results in
$E_{Th} \sim 90 \Delta$, and a fit to a quadratic behavior
of the form $a+b n^2$ leads to $b=2.48 \ 10^{-5}$,
while for the SVD unfolding $E_{Th^*}\sim 800 \Delta$.
Comparing with the typical spacing difference between
realizations for the box distribution which for $L=15$ equals to
$\delta \Delta = 5.17 \ 10^{-3}$, one
would estimates $E_{Th}= 0.44 \Delta^2/\delta \Delta \sim 85 \Delta$
and $b=(\delta \Delta / \Delta)^2 \sim 2.67 \ 10^{-5}$, both in good agreement
with the numerical results.
As explained above for $E_{Th^*}=r \Delta/2 k_{Th^*}$, and here
$r=3000$ and $k_{Th^*} \sim 2$,
resulting in $E_{Th^*}\sim 750\Delta$, again in good agreement with the results.

\section{Discussion}
\label{s5}

Much of the fascination with the SYK is connected to its chaotic behavior.
Since short time scales are associated with large energy scales, deviation
from GOE behavior of the spectra on large energy scales
indicate non-chaotic behavior on short
time scales. Estimating the time for which the chaotic behavior emerges is relevant
to the estimation of the scrambling time of the SYK models, which motivates
its application to studies of quantum gravity in black holes \cite{sachdev22}.
Ensemble averaging also plays an important role in the duality between SYK
and classical general relativity.

For self-averaging systems there is no difference between averaging over the ensemble
or averaging over a large single realization. As discussed, for finite size SYK
samples, there is a huge difference between the Thouless time for a realization
adjusted unfolding compared to the ensemble averaged unfolding, a difference which is
larger as the distribution of the interaction elements is more skewed.

Nevertheless,
it is relevant to extrapolate to infinite systems ($L \rightarrow \infty$) in order
to see whether this difference persists.
SYK has three time scales \cite{altland21}: (i) band structure time $t_B$
associated with the band width, $B$, (ii) Thouless time, $t_{Th}$ or $t_{Th^*}$,
on which much of the paper concentrated and
(iii) Heisenberg time, $t_{H}$. Since the band width depends linearly on $L$
\cite{cotler17}, $t_B=\hbar/B \sim \hbar/L$.
Thus, the Heisenberg time $t_H = \hbar/\Delta$ where
$\Delta \sim B/2^L$, is $t_H = \hbar\ 2^L/L$.
The Thouless time for the ensemble average is $t_{Th}=\hbar/E_{Th}$, where
$E_{Th} \sim \Delta L^2$ and 
resulting in $t_{Th}=\hbar\ 2^L/ L^3$. 
For the realization adjusted unfolding, $E_{Th^*} \sim 2^L \Delta B/k_{Th^*}$,
with $k_{Th^*} \sim O(1)$, thus, $t_{Th^*} \sim  k_{Th^*} \hbar/ B \sim k_{Th^*}/L$.
Hence, $t_H > t_{Th} \gg t_{Th^*} \geq t_B$, and while the realization adjusted
Thouless time $\lim_{L \rightarrow \infty}t_{Th^*} \rightarrow 0$, the
ensemble averaged Thouless time $\lim_{L \rightarrow \infty}t_{Th} \rightarrow \infty$.
The difference between
$t_{Th}$ and $t_{Th^*}$ increases as the distribution is more skewed
(Figs. \ref{fig2} and \ref{fig5}). 

This leads to the conclusion that 
one can not determine the behavior of the SYK model by ensemble
average for times shorter than $t_{Th}$ since shorter times are non-self-averaging. 
Moreover, these shorter times ($t_{Th^*} \sim t_B <t< t_{Th}$)
correspond to a parametically large span of times.
Therefore, when one wishes to study the transition from chaotic to localized
behavior in modified SYK models 
\cite{garcia18,garcia18a,nosaka18,garcia19,micklitz19,monteiro21,monteiro21a,jian17,jian17a,chen17,altland19,nandy22}
for which the number of independent random variables remain low, sample to sample fluctuations
are expected to remain important and non-self-averaging at short times should be considered.
In principal, although the transition occurs at long times nevertheless these effects could obscure the
transition point and influence the perceived nature of the metallic regime. 
This will be considered in future studies.




\begin{thebibliography}{99}

\bibitem{sachdev93} S. Sachdev and J. Ye, Phys. Rev. Lett. {\bf 70}, 3339 (1993).
  
\bibitem{kitaev15}
  A. Kitaev, Talks at the KITP on April 7th and May 27th
(2015).

\bibitem{sachdev15}  
S. Sachdev, Phys. Rev. X {\bf 5}, 041025 (2015).

\bibitem{maldacena16}
J. Maldacena and D. Stanford, Phys. Rev. D {\bf 94}, 106002 (2016).

\bibitem{maldacena99} J. Maldacena, International journal of theoretical physics {\bf 38}, 1113 (1999).
  
\bibitem{garcia16} A. M. Garc\'ia-Garc\'ia and J. J. M. Verbaarschot, Phys. Rev. D {\bf 94}, 126010 (2016).

\bibitem{you17}
Y-Z. You, A. W. W. Ludwig, and C. Xu,
Phys. Rev. B {\bf 95}, 115150 (2017).

\bibitem{li17}
T. Li, J. Liu, Yuan Xin and Y. Zhou,
J. High Energ. Phys. {\bf 06}, 111 (2017).

\bibitem{garcia18} A. M. Garc\'ia-Garc\'ia, Y. Jia, and J. J. M. Verbaarschot, Phys. Rev. D {\bf 97}, 106003 (2018).
  
\bibitem{garcia18a}
  A. M. Garc\'ia-Garc\'ia , B. Loureiro, A. Romero-Berm\'udez,
and M. Tezuka, Phys. Rev. Lett. {\bf 120}, 241603 (2018).

\bibitem{nosaka18}
  T. Nosaka, D. Rosa, and J. Yoon, The Thouless time,
  J. High Energ. Phys. {\bf 06}, 41 (2018).

\bibitem{garcia19}  A. M. Garc\'ia-Garc\'ia and M. Tezuka, Phys. Rev. B
  {\bf 99}, 054202 (2019).

\bibitem{micklitz19}
  T. Micklitz, F. Monteiro, A. Altland,
  Phys. Rev. Lett. {\bf 123}, 125702 (2019).

\bibitem{monteiro21}
F. Monteiro, M. Tezuka, A. Altland, D. A. Huse, T. Micklitz, 
Phys. Rev. Lett. {\bf 127}, 030601 (2021).

\bibitem{monteiro21a} F. Monteiro, T. Micklitz, M. Tezuka, and A. Altland,
Phys. Rev. Research {\bf 3}, 013023 (2021).

\bibitem{jian17}  
C.-M. Jian, Z. Bi, and C. Xu, Phys. Rev. B {\bf 96}, 115122
(2017).  
  
\bibitem{jian17a}
S.-K. Jian and H. Yao, Phys. Rev. Lett. {\bf 119}, 206602 (2017).

\bibitem{chen17}
 X. Chen, R. Fan, Y. Chen, H. Zhai, and P. Zhang,
Phys. Rev. Lett. {\bf 119}, 207603 (2017).

\bibitem{altland19}
  A. Altland, D. Bagrets, and A. Kamenev, Phys. Rev. Lett. {\bf 123}, 106601 (2019).


\bibitem{nandy22}
  D. K. Nandy,T. \~Cade\v{z}, B. Dietz, A. Andreanov, and D. Rosa, arXiv:2206.08599.
  
\bibitem{mehta91} M. L. Mehta, {\it Random matrices} (Acad. Press, New York,
1991), 2nd ed.

\bibitem{shklovskii93} B. Shklovskii, B. Shapiro, B. R. Sears, P. Lambrianides and H. B. Shore, Phys. Rev. B. {\bf 47}, 11487
  (1993).

\bibitem{ghur98}
T. Guhr, A. Muller-Groeling, H. A. Weidenmuller,  Phys. Rep. {\bf 299}, 190 (1998).

\bibitem{alhassid00}
Y. Alhassid,  Rev. Mod. Phys. {\bf 72}, 895 (2000).

\bibitem{mirlin00}
A.D. Mirlin, Phys. Rep. {\bf 326}, 259 (2000).

\bibitem{evers08}
R. Evers and A.D. Mirlin, Rev.  Mod. Phys. {\bf 80}, 1355 (2008).

\bibitem{altshuler86} B. Altshuler and B. Shklovskii, Sov. Phys. JETP [Zh. Eksp. Teor. Fiz. 91,220]
  {\bf 64}, 127 (1986).

\bibitem{sonner17} J. Sonner and M. Vielma,  J. High Energ. Phys. {\bf 11} 149  (2017). 

\bibitem{altland18} A. Altland and D. Bagrets, Nucl. Phys. B {\bf 930}, 45 (2018).
  
\bibitem{jia20} Y. Jia and J. J.M. Verbaarschot,  J. High Energ. Phys. {\bf 07} 193  (2020). 

\bibitem{aharony96}
A. Aharony and A.  Harris, Phys. Rev. Lett. {\bf 77}, 3700 (1996).
  
\bibitem{fossion13} R. Fossion, G. Torres-Vargas and J. C. L\'opez-Vieyra,
Phys. Rev. E, {\bf 88}, 060902(R) (2013).

\bibitem{torres17}
G. Torres-Vargas, R. Fossion, C. Tapia-Ignacio and J. C.
L\'opez-Vieyra, Phys. Rev. E, {\bf 96}, 012110 (2017).

\bibitem{torres18}
G. Torres-Vargas, J. A. M\'endez-Berm´udez, J. C. L\'opezVieyra and R. Fossion, Phys. Rev. E, {\bf 98}, 022110 (2018).
  

\bibitem{berkovits20}
  R. Berkovits, Phys. Rev. B {\bf 102}, 165140 (2020).

\bibitem{berkovits21}
  R. Berkovits, Phys. Rev. B {\bf 104}, 054207 (2021).
  
\bibitem{berkovits22}
  R. Berkovits, Phys. Rev. B {\bf 105}, 104203 (2022).
  
\bibitem{rao22} W.-J. Rao, Phys. Rev. B {\bf 105}, 054207 (2022).
  
\bibitem{khaymovich20}
I. M. Khaymovich, V. E. Kravtsov, B. L. Altshuler, and L. B. Ioffe,
Phys. Rev. Res. {\bf 2}, 043346 (2020).

\bibitem{atas13} Y. Y. Atas, E. Bogomolny, O. Giraud, and G. Roux, Phys. Rev.
Lett. {\bf  110}, 084101 (2013).

\bibitem{sachdev22} S. Sachdev, arXiv:2205.02285.

\bibitem{altland21} A. Altland, D. Bagrets, P. Nayak, J. Sonner, and M. Vielma,
Phys. Rev. Res. {\bf 3}, 033259 (2021).

\bibitem{cotler17}
J. Cotler, G. Gur-Ari, M. Hanada, J. Polchinski, P. Saad, S. Shenker, D. Stanford, A. Streicher, and M. Tezuka,
J. High Energ. Phys. {\bf 05}, 118 (2017).

\end{thebibliography}
\end{document}